\documentclass[a4paper,11pt]{article}
\usepackage{jheppub}

\usepackage{amsmath}
\usepackage{amsfonts}
\usepackage{amssymb}
\usepackage{graphicx}
\usepackage{wrapfig}
\usepackage{braket}
\usepackage{hyperref}
\usepackage{amsthm,verbatim,cancel}
\usepackage{color}
\usepackage{caption}
\usepackage{subcaption}
\newcommand{\pd}{\partial}
\newcommand{\bea}{\begin{eqnarray}}
\newcommand{\eea}{\end{eqnarray}}
\newcommand{\nn}{\nonumber\\}

\DeclareMathOperator{\tr}{tr}

\begin{document}

\title{A note on entanglement edge modes in Chern Simons theory }

\author[a,b]{Gabriel Wong}
\emailAdd{gabrielwon@gmail.com}
\affiliation[a]{Department of Physics, University of Virginia, Charlottesville, VA 22901}
\affiliation[b]{Perimeter Institute for Theoretical Physics, 31 Caroline St, N. Waterloo, ON N2L 2Y5, Canada. }

\abstract{We elaborate on the extended Hilbert space factorization of Chern Simons theory and show how this arises naturally from a proper regularization of the entangling surface in the Euclidean path integral.  The regularization amounts to stretching the entangling surface into a co-dimension one surface which hosts edge modes of the Chern Simons theory when quantized on a spatial subregion.  The factorized state is a regularized Ishibashi state and reproduces the well known topological entanglement entropies.  We illustrate how the same factorization arises from the gluing of two spatial subregions via the entangling product defined by Donnelly and Freidel \cite{Donnelly:2016auv}.  }

\maketitle
\section{Introduction}
In recent years, quantum information theory has found many important applications in high energy and condensed matter physics.  The central idea that connects these different fields is quantum entanglement, which refers to the way different sub-systems are correlated with one another.  In high energy physics, recent work in AdS/CFT \cite{VanRaamsdonk:2010pw}\cite{Swingle:2014uza}\cite{Faulkner:2013ica} suggests that quantum entanglement is responsible for the emergence of spacetime.  Meanwhile, in condensed matter physics, entanglement has emerged as an important diagnostic for topological phases \cite{Li:2008kda}.  
To define entanglement between a spatial subsystem $A$ and it's complement $ B$ we must factorize the Hilbert space 
\bea \label{fact}
H = H_{A} \otimes H_{B} 
\eea
The factorization allows us to define the reduced density matrix $\rho_{A}$ from a state $\rho$ by tracing out the complement $B$
\bea \label{t}
\rho_{A}= \tr_{B}\rho 
\eea
We define the modular Hamiltonian $\mathcal{H}_{A}$  by 
\bea \label {mod} 
\rho_{A} = \frac{e^{-\mathcal{H}_{A}} }{Z_{A}}
\eea
where $Z_{A}$ is a normalization constant.  $\mathcal{H}_{A}$
acts as an effective Hamiltonian in the region $A$.
In a continuum theory, the Hilbert space factorization \eqref{fact}  requires a UV regularization that separates region $A$ and $B$.  Among other things, this is needed to  define the partial trace \eqref{t} in the continuum.  In principle such a regularization is provided by a lattice cutoff.  However, in a gauge theory, the Gauss law constraint poses an obstruction to the naive factorization even on the lattice.  One way to resolve this problem is to relax Gauss's law so that it can be violated at the boundary of $A$ and $B$.  Formally this requires embedding $H$ into an \emph{extended} Hilbert Space \cite{Donnelly:2014gva}
\bea
H \subset H_{A}\otimes H_{B},
\eea
where $H_{A}$ and $H_{B}$ are allowed to contain surface charges at their respective boundaries that transform non-trivially under the boundary gauge group.  These additional degrees of freedom, which we refer to as ``entanglement edge modes'', represent the minimal extension needed to construct a factorizable Hilbert space from $H$.  The physical Hilbert space is contained inside a gauge invariant subspace of $H_{A} \otimes H_{B}$: within the enlarged Hilbert space, one can trace over $B$ and define the entanglement entropy as 
\bea
S_{A}=- \tr \rho_{A} \log \rho_{A} 
\eea
In the physical, gauge invariant state, the Gauss law requires a matching of the electric field across the boundaries which induces entanglement between the edge modes on $\pd A$ and $\pd B$.  For example, in a lattice gauge theory where each link supports a Hilbert space spanned by the representation matrix elements of the gauge group, the entanglement entropy (EE) takes the general form \cite{Donnelly2011},\cite{Buividovich:2008yv}:
\bea
S_{A} = - \sum_{R_{\pd}} P(R_{\pd}) \log P(R_{\pd}) + \sum_{R_{\pd}} P(R_{\pd }) \sum_{e \in \pd} \log \dim R_{e} + \text{bulk EE}
\eea
where $R_{\pd}$ denotes the set of representations labeling the links on the boundary of region $A$, $e$ denotes the individual edges on the boundary and $\dim R_{e}$ is the dimension of the representation $R_{e}$ associated with an edge $e$.  The first two terms come from the edge modes: the first is the Shannon entropy associated with the probability distribution $P(R_{\pd}) $ of representations on the boundary and the second term counts the internal states transforming under the non-abelian gauge group.  It may seem strange to include the entropy due to edge modes, which  are gauge variant degrees of freedom added for the sake of factorizing the Hilbert space.  However, recent work have shown that these terms have a real significance.  For example in $U(1)$ Maxwell theory, the Shannon-entropy of the edge modes is necessary to reconcile the replica trick and thermal calculation of entanglement entropy \cite{Donnelly:2014fua,Donnelly:2015hxa}, and in topological gauge theories like Chern Simons theory, the $\log \dim R$ term is responsible for the topological entanglement entropy, which is an important diagnostic for topological phases.   

In \cite{Donnelly:2016auv},the authors developed a classical phase space analog of extended Hilbert space for Yang- Mills and Einstein gravity and showed how the edge degrees of freedom are needed to define the theories in a subregion in a gauge invariant way. 
So far, few examples of the extended Hilbert space have been worked out in the continuum with the exception of two-dimensional Yang-Mills and U(1) Maxwell theory.   In this work, we show how to derive the extended Hilbert space for Chern Simons theory in the frame work of \cite{Donnelly:2016auv} and explain how this naturally arises from a proper regularization of the Euclidean path integral which prepares the quantum state in question.  
Many of the results in this work have appeared in the literature in various guises.  The appearance of physical edge modes in the entanglement spectrum of topological phases was first discussed in \cite{Li:2008kda},\cite{Swingle:2011hu},\cite{qi2012general} and the computation of topological entanglement entropy in terms of left-right entanglement of boundary states was done in 
\cite{Das:2015oha}. The reference \cite{Wen:2016snr}, noted a problem with the approach of \cite{qi2012general}, where the reduced density matrix was obtained by a quantum quench in which region $A$ and $B$ are disconnected suddenly.  The initial condition for such a quench is given by a conformally invariant boundary state $\ket{B}$, satisfying 
\bea
(L_{n}-\bar{L}_{-n}) \ket{B}=0.
\eea
A general solution to this equation is given by a linear combination of Ishibashi states, which we will discuss in more detail below.   Modular invariance dictates that $ \ket{B}$ satisfies the Cardy condition, which implies that $ \ket{B}$ is a particular linear combination of the Ishibashi states, with the coefficients given by elements of the modular S-matrix.   Unfortunately, the state $\ket{B}$ does not reproduce the known entanglement entropies in Chern Simons theory, and the authors of \cite{Wen:2016snr} showed how to obtain the correct entropies by relaxing the Cardy condition.  

In this work we offer a different perspective that justifies the methods of \cite{Wen:2016snr}. We carry out two simple derivations of the Chern Simons extended Hilbert space that lead to an explicit expression for embedding of the Chern Simons wave-functionals into the extended Hilbert space.  In section \ref{edge modes}, we show how this embedding naturally arises from a careful UV regularization of the entangling surface in the path integral description of the reduced density matrix $\rho_{A}$.  This section is essentially an application of the old arguments of Unruh \cite{Unruh:1983ac} regarding the entanglement of the Minkowski vacuum to holographic topological quantum field theories.  In section \ref{eprod} we arrive at the same results by implementing the ``entangling product" defined in \cite{Donnelly:2016auv}, which amounts to glueing $H_{A}$ and $H_{B}$ into the ``bulk'' Hilbert space  $H$.   In section 4 we discuss the calculation of entanglement entropy directly from the reduced density matrix.  In the conclusion, we will end with some speculations about the description of the entanglement edge modes in the string theory dual to Chern Simons theory, which served as the original motivation for this work.  

While this work was being competed, we became aware of  \cite{Fliss:2017wop} which takes a similar perspective on the edge modes of Chern Simons theory.  

\section{ Edge modes from the Euclidean Path integral }
\label{edge modes}

\subsection{Path integral definition of the reduced density matrix}
In the ground state of a continuum QFT,  the matrix elements of the (un-normalized) reduced density matrix $\tilde{\rho}_{A}$ can be represented by a path integral with a cut along region $A$
\bea\label{pt}
\braket{ \phi_{A} |\tilde{\rho}_{A}| \phi'_{A}}=  \int D[\phi] e^{-S[\phi]} \delta (\phi_{A^{+}} -  \phi_{A}) \delta (\phi_{A^{-}} -\phi'_{A} ) 
\eea
where $A^{\pm}$ refers to the upper and lower edge of the cut.  While this representation is adequate for the computing the entanglement entropy via the replica trick, it obscures the factorization of the Hilbert space.   What should be done with the degrees of freedom living on the entangling surface $\pd A$, which naively belongs to both region A and B? 
As noted in  recent works \cite{Cardy:2016fqc} \cite{Donnelly:2016jet}, the resolution is to remove a small tubular neighborhood of the entangling surface, resulting in a stretched co-dimension 1 space-time boundary.  For certain backgrounds and choice of $A$ and $B$,  we can choose the angular coordinate $\theta$ around the entangling surface as the  Euclidean time coordinate.   In such cases we can write the  trace of $\tilde{\rho}_{A} $ as a path ordered exponential in $\theta$
\bea
Z_{A}= \tr_{A} \tilde{\rho}_{A} = \text{P} \exp \left( - \int_{0}^{2\pi} K_{A}(\theta) \, d\theta \right)
\eea
where $K_{A}$ generates translations in $\theta$.  In a topological theory $K_{A}$ is a conserved charge independent of $\theta$ so this reduces to 
$Z_{A}= e^{-2 \pi K_{A}}$, which can be interpreted as a thermal partition function at temperature $2 \pi $.  The Modular Hamiltonian $\mathcal{H}_{A} = 2 \pi K_{A}$ is thus a local Hamiltonian on $A$, and the entanglement entropy is identified with the thermal entropy of $Z_{A}$.   Due to the presence of the stretched entangling surface, $\mathcal{H}_{A}$ will in general have boundary terms describing the edge degrees of freedom.  In the next section we will derive the Modular Hamiltonian for a disk in Chern Simons gauge theory and show how these edge modes are entangled.


\subsection{ The extended Hilbert space of Chern Simons theory } 
The Chern Simons Hilbert space on a closed, compact Riemann surface $\Sigma$ is equivalent to the space of conformal blocks of a chiral WZW model on $\Sigma$ \cite{Witten:1988hf}.   On the other hand, when we make a spatial decomposition  
\bea \label{P}
\Sigma= A \cup B
\eea
the Chern Simons Hilbert spaces on $A$ and $B$ are those of the edge chiral WZW model on $\pd A$ and $\pd B$.  By preparing states in $H_{\Sigma}$ with the Euclidean path integral we will provide an explicit formula for their embedding into the extended Hilbert space 
\bea \label{factor}
H_{\Sigma} \subset H_{A} \otimes  H_{B} 
\eea 
where $H_{A}$ and $H_{B}$ are the chiral and anti chiral Hilbert spaces living entirely on the edge.
 
The simplest setting in which we can  realize \eqref{factor} is for the vacuum state on  $\Sigma=S^{2}$ and a disk-like region $A$.  This state $\ket{\psi}$ is prepared by the Euclidean path integral on a solid ball $B^3$ with the spatial $S^2$ as its boundary.  Just as we did for the reduced density matrix, we will regulate this geometry by removing a semi-tubular neighborhood of the circle $\pd A$ of radial size $\epsilon$.   We can then slice the path integral using an angular time coordinate $\theta$  that encircles the entangling surface.  The advantage of this time slicing is that the (un-normalized)  wave functional $\braket{\phi_{A},\phi_{B} | \psi }$ can be viewed as an amplitude between states living on $A$ and $B$        \cite{Jacobson:2012ei}:  
\bea  \label{CJ}
\braket{\phi_{A},\phi_{B} | \psi } = \braket{\phi_{A} |  e^{- \pi K_{A}} \mathcal{J} |{ \phi}_{B}}
\eea
where $K_{A}$ generates rotations around the entangling surface and $\mathcal{J} $ is an anti-linear operator that implements a CPT transformation. 
 \begin{align}
\mathcal{J} : &H_{B} \rightarrow H_{A} 
\end{align}
 The parity transformation is needed because the $\pi$ rotation on the RHS of \eqref{CJ} requires that $\ket{\phi_{A}}$ and $\mathcal{J}  \ket{\phi_{A}}$ have opposite orientations, the C transformation accounts for the bra $\rightarrow$ ket mapping  $\bra{\phi_{A}} \rightarrow \ket{\phi_{A}}$, and the $T$ transformation accounts for the fact that $\theta $ and $t$ are oppositely oriented on region $B$ \cite{Jacobson:1988qt}.   
\begin{figure}
\centering 
\includegraphics[scale=.3]{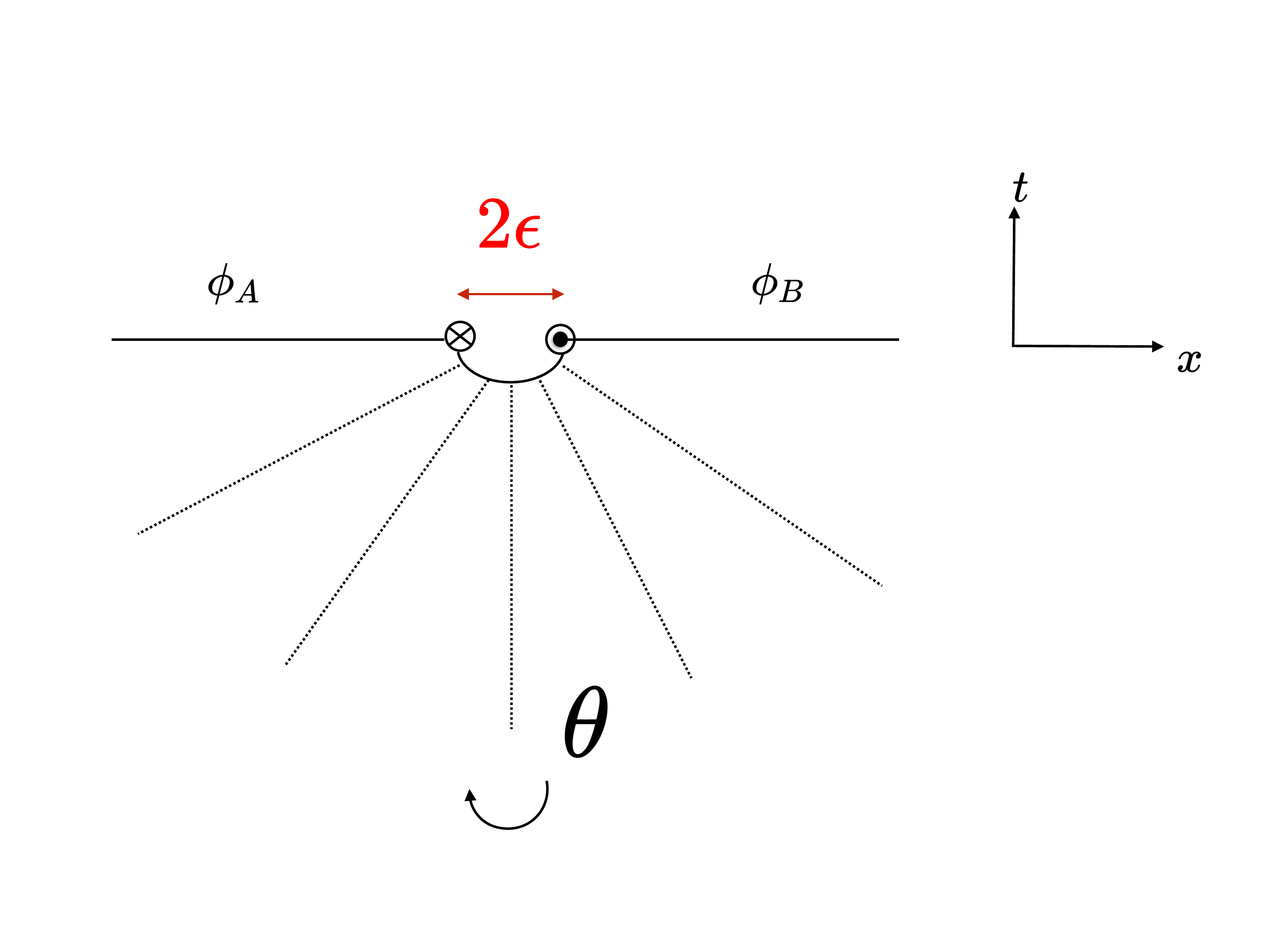}
\caption{The wavefunctional prepared by the Euclidean path integral can be sliced in angular time $\theta$, provided we remove an $\epsilon$ neighborhood of the entangling surface.  When viewed as the initial and final time slices in $\theta$,  region $A$ and $B$ are assigned  opposite orientations relative to the ambient space $A\cup B$.  Thus they support the Hilbert space of a chiral and anti-chiral edge CFT, which transforms non-trivially under the boundary gauge group.  In addition to parity, switching from $t$ to $\theta$ also involves a time reversal because they have opposite orientations on region $B$ at the $t=0$ time slice.  Sliced in angular time, the Euclidean path integral prepares a Thermofield double state in which the left right entanglement of these CFT's leads to singlet state under a diagonal action of the boundary gauge group on $\pd A$ and $\pd B$  }
\end{figure}
The operator $K_{A}$ evaluated at $\theta=0$ generates infinitesimal translations in the $t=x^{0}$ direction and is therefore the integral of the physical energy density $T_{00}$ weighted by a function $f(x)$
\bea \label{H}
K_{A} =  \int_{A}d^{D-1}x   \,\, f(x) \, T_{00} 
\eea
In the bulk of $A$, $T_{00}$ is zero.  As noted earlier, quantizing Chern Simons theory on $A$ leads to a chiral CFT Hilbert space living on the edge.  Thus the energy density is a delta function supported on the boundary $\pd A$, which we take to be the ``stretched'' surface at distance $\epsilon$ away from the entangling surface.
\bea \label{T00}
 T_{00}= \delta (\pd A)   \,\, T_{\text{CFT}}.
\eea 
In a neighborhood of the entangling surface we can choose local coordinates $(r,\theta, y^{i})$, with $r$ a radial coordinate away from the entangling surface and $y^{i}$ parametrizing the surface itself.  In this neighborhood, 
$f(x)\sim r$, so that $K_{A}$ takes the form of a (Euclidean) Rindler Hamiltonian.   Due to the delta function in \eqref{T00} we can relate $K_{A}$ to the Hamiltonian of the chiral edge CFT:
 \begin{align}
  K_{A} &= \epsilon  \int_{\pd A}  T_{\text{CFT}} = \epsilon H_{\text{CFT}} 
\end{align}    
For a circular boundary of length $l$ we can express the Hamiltonian in terms of the right moving Virasoro generator $L_{0}$ and the central charge $c$.  
\bea
H_{\text{CFT}}&=\frac{2 \pi } {l} (L_{0}  - \frac{c}{24})  
\eea
Expanding in a complete set of eigenstates of the Modular Hamiltonian in $A$ and $B$ in \eqref{CJ} and using the anti linearity of $\mathcal{J} $ leads to a representation of  $\ket{\Psi}$ as a \emph{thermofield double state} of the boundary CFT's.
\begin{align}
\braket{\phi_{A} |  e^{- \pi K_{A}} \mathcal{J}  |{\phi}_{B}} &=\sum_{m,n} \braket{\phi_{A}|n_{A}}\braket{n_{A}|e^{- \pi K_{A}} \mathcal{J} ( | m_{B}}\braket{m_{B}|\phi_{B}})\nn
&=\sum_{n,m} \braket{\phi_{A}|n_{A}}\braket{n_{A}| \mathcal{J}|m_{B}}    \braket{\phi_{B}|m_{B} } e^{- \pi \epsilon E_{n}}  \nn
&= \bra{\phi_{A} \phi_{B}}\sum_{m} e^{-\pi \epsilon E_{n}} \ket{\bar{m}_{A}}\ket{m_{B}}
 \end{align}
 where we used the anti-linearity of $\mathcal{J} $ in the second line and denoted $ \ket{\bar{m}_{A}} = \mathcal{J}\ket{m_{B}}$ in the third line \cite{Jacobson:2012ei}.  Adding the normalization factor $Z_{A}$, we thus arrive at the extended hilbert space factorization of the ``bulk" state $\ket{\psi}$
\bea\label{TFD}
\ket{\psi}= \frac{1}{Z_{A}} \sum_{n} e^{-  \epsilon  \pi E_{n}} \ket{\bar{n}_{A} }\otimes \ket{n_{B}}
\eea
where $\ket{n_{B}}$ and $ \ket{\bar{n}_{A} } $ are CPT conjugate states, corresponding to the right and left moving energy eigenstates of the edge CFT's, and $E_{n}$ are the right moving eigenvalues.  Thus the bulk entanglement between the two disks are given entirely by the left-right entanglement of the edge CFT, as noted previously in \cite{qi2012general}, \cite{Das:2015oha}.  

Equation \eqref{TFD} gives an explicit embedding of the state $\ket {\psi} $ into the extended Hilbert space.  The state $ \ket{I} =\sum_{n}\ket{\bar{n}_{A}} \otimes \ket{n_{B}}$ is an Ishibashi state \cite{Ishibashi:1988kg}, which solves the conformally invariant boundary condition 
\bea\label{cbc}
(L_{m}-\bar{L}_{-m})\ket{I}=0
\eea
The thermo-field double state in \eqref{TFD} is a regularized Ishibashi state that has been rendered normalizable by evolving it with the CFT  Hamiltonian for a small time $\epsilon$.
We will see the relevance of \eqref{cbc} in the next section. 

From \eqref{TFD} we can identify the $H_{\text{CFT}}$ with the modular Hamiltonian.  The reduced density matrix on $A$ is just the thermal density matrix at  inverse temperature $2 \pi \epsilon$: 
\bea
\rho_{A} = \frac{e^{ -2 \pi \epsilon H_{\text{CFT}}}}{Z_{A}}
\eea
The normalization $Z_{A}$  is given by the finite temperature partition function of CFT. 
\bea
Z_{A}  = \tr_{A} e^{ -2 \pi  \epsilon H_{\text{CFT}}}
\eea
Thus we see the edge modes partition function arise directly via the regularization of the entangling surface. 
For generic regions $A$ and $B$, we may not be able to find a globally well defined coordinate $\theta$ on the three manifold $M$  that evolves one region onto the other. \footnote{However, after regularizing, we can always find such a coordinate locally near the entangling surface that evolves $\pd A$ into $ \pd B$.   Since $T_{00}$ vanishes in the bulk, this may be sufficient to generalize the path integral argument to other regions. }  In the next section we will provide another argument for the extended Hilbert space factorization , based on the requirements of gauge invariance in the presence of the entangling surface, which generalizes more readily to an arbitrary region A.

\section{ The entangling product for Chern Simons theory}
\label{eprod}
In the previous section we applied the extended Hilbert space construction \cite{Donnelly:2014gva}, \cite{Donnelly:2014fua} in which a gauge invariant bulk state $\ket{\psi}\in H_{\Sigma} $ was factorized by  embedding into a larger, non-gauge invariant Hilbert space $ H_{A} \otimes H_{B}$ consisting of left and right moving edge modes.   Now we consider this procedure in the opposite direction:  starting with the Hilbert spaces $H_{A}$ and $H_{B}$, how do we glue them together to produce a gauge invariant subspace containing the bulk state $ \ket{\psi} $?   This ``entangling product"  was constructed for pure Yang Mills (and Einstein gravity) at the classical level and implemented quantum mechanically in the case of Yang Mills in 1+1 D in \cite{Donnelly:2016auv}. More recently, the classical phase space description of the entangling product  was applied to abelian Chern Simons theory in \cite{Geiller:2017xad},\cite{Fliss:2017wop}.   Below we will implement the quantum mechanical  entangling product in Chern Simons theory. 
\subsection{Hilbert space on a disk}
To begin, we recall how imposing gauge invariance in the presence of a boundary leads to new boundary degrees of freedom that transform under a boundary symmetry group.   The Chern Simons action 
\bea
S_{0}= \frac{k}{4 \pi } \int _{M}  \, \tr( A \wedge dA + \frac{2}{3} A\wedge A \wedge A )
\eea
on a manifold $M$ is not gauge invariant when $\pd M$ has a boundary.  This is because the on-shell variation gives a boundary term: 
\bea\label{dS} 
\delta S_{0}\big|_{\text{on-shell}}  = \int_{\pd M}  \, \tr(\delta A \wedge A) ,
\eea
One way to restore gauge invariance is to restrict gauge transformations to be trivial on the boundary.  This means that the would-be pure gauge degrees of freedom living on $\pd M$ are promoted to dynamical degrees of freedom.  These edge modes transform non trivially under the boundary gauge group, which is promoted to a physical symmetry.  

To be explicit, consider the the manifold $M = D \times S^{1}$ with $D$ a disk. We have compactified the time dimension, which we will interpret as the angular direction $\theta$ around the entangling surface.  Quantization of Chern Simons on a disk was worked out in \cite{Elitzur:1989nr},\cite{Witten:1988hf}.  Following  \cite{Elitzur:1989nr}, we choose the boundary condition $A_{0}=0$ to set the boundary term \eqref{dS} to zero, thus obtaining a well defined variational principle for the bulk equations of motion. 
The gauss law constraint $\frac{\delta S}{\delta A_{0}}=0$  restricts to flat connections on $D$, which are of the form.
\bea \label{A}
A= - \tilde{d} U U^{-1}
\eea
where $\tilde{d}$ is the spatial exterior derivative on $D$ , and $U \in G$ is an element of the gauge group $G$.  Note that this is only pure gauge in the bulk, since gauge transformations are required to be trivial on the boundary. 
Inserting \eqref{A} back into the Chern Simons action gives the chiral WZW action
\bea
S_{\text{WZW}} = \frac{k}{4 \pi} \int_{\pd D \times R} \tr (U^{-1} \pd_{\phi}UU^{-1} \pd_{t}U) d\phi d t + \frac{k}{12\pi} \int_{D\times R} \tr [(U^{-1} dU)^3]   
\eea
This action only depends on the boundary values\footnote{The second term involves an integration over the bulk but for integer $k$ different bulk extensions of $U$ differ by integer multiples of $2\pi i$ , therefore giving the same exponential weighting in the path integral.} of $U$.  Gauge fixing and dividing by the volume of the gauge group in the bulk then leads to a boundary theory with the path integral
\bea
Z_{\text{bdry}}=\int d[U] e^{-S_{\text{WZW}}}
\eea
where $U : \pd D \times R \rightarrow G$.     Thus Chern Simons theory on a disk is equivalent to a chiral WZW model living on the edge.  

As alluded to earlier,  the boundary symmetry group consist of the gauge transformation restricted to the boundary.  In order to preserve the boundary condition $A_{0}=0$ these are also required to be time independent.  Thus the boundary symmetry group is the loop group $LG$, whose elements  
\bea
g: S^{1}  \rightarrow  G,
\eea
are maps from the boundary circle to the gauge group. 
Explicitly, the loop group elements acts on the WZW model via
\bea
U \rightarrow g (\phi)  U.
\eea
The Hilbert space thus furnishes a representation of the loop group, or equivalently the current algebra of the gauge group G.   The currents generating this algebra are just the boundary values of the gauge field 
\bea
J^{a}(z)= A^{a}_{z}= (\pd_{z} U^{-1}\, U)^{a}\nn
a=1,..\dim G
\eea
where $z$ is a holomorphic coordinate on the boundary spacetime, and $a$ is a group index. 

For example, in the abelian case with gauge group $G= U(1) $, $U=e^{i \theta }$ and the WZW model reduces to a chiral, compact boson $\theta$ with radius $\sim\frac{1}{k}$.  The modes of the boson current $J(z)= \pd_{z} \theta$ satisfy the infinite dimensional current algebra at level k:
\begin{align}
[J_{n},J_{m}] = k \,  n \delta_{n+m}
\end{align}
Up to the $k$ dependent normalization, this is just an infinite set of harmonic oscillators, with the $n>0$ currents acting as annihilation and $n<0$ acting as creation operators.   A representation of this algebra is obtained by applying $J_{-n}$ to the highest weight state $\ket{0}$ which is annihilated by $J_{n}$ for $ n>0 $.  
When $G$ is non-abelian, the level $k$ current algebra is
\bea\label{kacs}
[J^{a}_{m},J^{b}_{n}] =i f^{abc}J^{c}_{m+n} +k \, m \delta^{ab}  \delta_{n+m},
\eea
where $ f^{abc}$ are the structure constants of $G$.  In this case the zero modes $J_{0}^{a}$ generate the finite dimensional Lie algebra of $G$ and the highest weight states $\ket{r}$ transforms in an irreducible representation $r$ of G.   Only a subset of representations $r$ of $G$ admit the infinite dimensional generalization \eqref{kacs}, and each corresponds to a primary field.  For example, for $G=SU(2)$, the ground states transform in the usual $2r +1$ dimensional representations with basis elements:
\bea
\ket{r, m} \,\ ,  |m|\leq r
\eea
and the Hilbert space is created from the highest weight state $\ket{ r, r}$ via the ladder operators $J^{a}_{-n}$ and $J_{0}^{1} - i J^{2}_{0}$.   The allowed ``integrable'' representations corresponds to half-integers $r$ satsifying $ 0\leq r < k$.  From the point of view of the bulk Chern Simons theory, states in different representations are prepared by a path integral Wilson line inserted inside  $D\times R$ in the representation $r$\cite{Witten:1988hf} \cite{Elitzur:1989nr}. 

\subsection{Gluing of Hilbert spaces}

Now let us return to the original question of how to glue together two Hilbert spaces $H_{A}$ and $H_{B}$.  We take $A=D$ and $B=\bar{D}$ to be two oppositely oriented disks, which we wish to glue into a sphere $S^{2}=D \cup \bar{D} $.  $H_{A}$ and $H_{B}$ provides a representation of the chiral currents $ J^{a}(z)$  anti-chiral currents $\bar{J}^{a} (\bar{z} )$ respectively.  
Since the total Hilbert space $H_{S^2} $ is gauge invariant, whereas the tensor product $H_{A} \otimes H_{B}$ transforms non-trivially under $G$, we must restrict to a gauge invariant subspace.  This subspace is denoted by the entangling product 
\bea
H_{S^{2}} =H_{A} \otimes_{G} H_{B} 
\eea
defined as a quotient of the tensor product $H_{A} \otimes H_{B} $ by the simultaneous action of the boundary symmetry group on  $\pd A$ and $\pd B$.  In other words, a state  $\ket{\Psi} \in H_{A} \otimes_{G} H_{B} $ is a singlet under the diagonal action of the loop group on the two edges of opposite chirality.   Such a state is invariant under the current algebra and satisfies the constraint \cite{Ishibashi:1988kg} 
\bea \label{JJ}
( \bold{1_{A}} \otimes J^{a}_{n}   + \bar{J}^{a}_{-n} \otimes  \bold{1_{B}} )\ket{\Psi}=0
\eea
For each representation of the current algebra corresponding to a primary with weight $r$, there is a solution to this equation  given by the an Ishibashi state \cite{Ishibashi:1988kg}   
\begin{align}\label{Ishi}
\ket{I, r} &=  \sum_{n} \ket{r,\bar{ n}}_{A} \otimes \ket{r ,n}_{B},
\end{align}
where $\ket{r, n}_{B}$ is a basis in the $r$ representation of the chiral current algebra, and the CPT conjugate $\ket{r,\bar{n}}_{A}= \mathcal{J} \ket{r, n}_{B}$ is the anti chiral counterpart. 
We can see that this is a solution by taking an arbitrary state $\ket{\bar{a}}\ket{b}$ and computing the  the overlap 
\begin{align}
\bra{\bar{a}}\otimes\bra{b} (\bar{J}^{c}_{m} + J^{c}_{-m} ) \sum_{n} \ket{\bar{ n}} \otimes \ket{n}&= \sum_{n} \braket{\bar{a}|\bar{J}^{c}_{m}|\bar{n}}  \braket{b|n} + \braket{\bar{a}|\bar{n}} \braket{b|J^{c}_{-m}|n}\nn
&= \sum_{n}  \braket{ b | n}  \braket{a|\mathcal{J}^{\dagger} \bar{J}^{c}_{m} \mathcal{J}|n}^{*} + \braket{\mathcal{J} a| \mathcal{J} n} \braket{b| J^{c}_{-m}|n}\nn
&= \sum_{n}  - \braket{b|n}  \braket{a|J^{c}_{m}|n}^{*} +\braket{b| J^{c}_{-m}|n}\braket{n|a} \nn
&= \sum_{n}  - \braket{b|n}  \braket{n|J^{c}_{-m}|a}+ \braket{b| J^{c}_{-m}|n}\braket{n|a} =0
\end{align} 
Here we have made use of the identities $\braket{\mathcal{J} a|\mathcal{J}b}= \braket{a|b}^{*}$ and $\braket{a|\mathcal{J}b}=\braket{\mathcal{J}^{\dagger} a|b}^{*}$ due to the anti-linearity of $\mathcal{J}$.
For the vacuum representation, normalizing the state \eqref{Ishi} by applying the CFT evolution operator gives the same result obtained in the previous section by path integral methods.  More generally, the choice of representation is determined by the bulk state $\ket{ \Psi_{r}}$, which is prepared by the path integral on a solid ball $B^3$ with a Wilson line in the $r$ representation inserted.  The endpoints of the Wilson line corresponding to anyonic charges $r$ and $ \bar{r}$ are inserted inside  $D$ and $\bar{D}$ of the bounding sphere $S^{2} =\pd B^3$.   
Accounting for the normalization, such a state factorizes as
\bea
\ket{ \Psi_{r}}= \frac{1}{Z_{A}} \sum_{n}  e^{- \pi \epsilon E_{n}} \ket{r, \bar{ n}}_{A} \otimes \ket{r ,n}_{B}
\eea
The constraint \eqref{JJ}, also implies the conformally invariant boundary condition \eqref{cbc}.  This is because the the generators of the current algebra is directly related to the Virasoro generators via  the Sugawara construction \cite{Ginsparg:1988ui}: 
\bea
L_{n} \sim \sum_{m} :J^{a}_{m+n}J^{a}_{-m}:
\eea

\section{Entanglement entropy}

The entanglement entropy can be organized in a thermal form as a sum of a modular energy and a free energy term: 
 \bea
 S_{A} = - \tr \rho_{A} \log \rho{A}= \tr_{A}(\rho_{A} H_{A})- \log Z_{A}
 \eea
When $A$ is a disk with charge $r$ inserted, we have 
\bea
S=   2 \pi \epsilon \tr_{A}( \rho_{A} H_{\text{CFT}} )+ \log ( \chi_{r}( e^{-\frac{ 2 \pi \epsilon}{l}}))
 \eea
 The first term vanishes as $ \epsilon \rightarrow 0$ , so the entropy comes entirely from the free energy, which we have written in terms of the Virasoro character  $\chi_{r}$ .  The series is badly behaved as $ \epsilon \rightarrow 0$ due to the infinite temperature limit, but can be computed by applying a modular transform \cite{Wen:2016snr}. 
\bea
\chi_{r}( e^{- \frac{ 2 \pi \epsilon}{l}})= \sum_{s} S_{r}^{s} \, \chi_{s} (e^{- \frac{2 \pi l }{ \epsilon}})  \rightarrow  S_{r 0} \,\chi_{0} (e^{- \frac{ 2 \pi l }{ \epsilon}})
\eea 
where in the last line we have retained the dominant term in the $\epsilon \rightarrow 0 $ limit.  In the limit, the partition function  $\chi_{0}$ is dominated by the ground state with Casimir energy $E_{0} = \frac{ -\pi c}{6}$, giving  $\chi_{0} \sim e^{ \frac{2\pi l}{\epsilon}  \frac{ \pi c}{6}}$
This gives the entropy
\begin{align}
S_{A}&=\frac{ \pi^{2} c  l }{ 3  \epsilon} + \log (S_{r0})
\end{align}
The first term is the standard area law with explicit dependence on the regulator, and the second is the well known topological entanglement entropy.   Since $S_{r0}=\frac{d_{r}}{\mathcal{D}}$,  where $d_{r}$ is the quantum dimension for the represention $r$ and $\mathcal{D}$ the total quantum dimension, we can write the topological term as
\bea
S_{\text{top}} = - \log \mathcal{D} + \log d_{r} 
\eea
We can interpret the extra term $\log d_{r}$ relative to the vacuum state as being due to the additional entanglement between anyonic charges.   In the replica trick calculation, this arises because the path integral representation for $\tr_{A} \rho_{A}^n$ computes the expectation value of Wilson lines that have been inserted in the bulk \cite{Lewkowycz:2013laa}.  

\section{Conclusion}

In this work, we have provided two explicit derivations v of the extended Hilbert space factorization of Chern Simons theory.  Both are universal and applicable to holographic TQFT's which host edge modes in the presence of a boundary.  The essential ingredient involved in both derivations is the regularization of the entangling surface, leading into a codimension one boundary.  This is reminiscent of the stretch horizon in the study of black hole physics.  The reduced density matrix describe a thermal ensemble of the edge theory and the entanglement entropy is given by the corresponding thermal entropy.   The factorization of the bulk state into a maximally entangled Ishibashi state of two boundaries can interpreted as a gluing of two spacetimes along their edges.   It would be interesting to apply this gluing in three dimensional gravity, which can also be formulated as a Chern Simons theory.   

Chern Simons theory has no local degrees of freedom.   Yet, the extended Hilbert space construction provides a precise definition of entanglement between spatial regions.   In the presence of Wilson lines, we can think of the entanglement entropy ( relative to the vacuum with no insertions ) as being due to the cutting of the Wilson line by the entangling surface.  String theory is another important example of a theory with non-local degrees of freedom.  It is tempting to think that the fundamental string also induces entanglement across spatial regions when cut by an entangling surface.

Unfortunately, very little is known about the entanglement structure of string theory.   It was conjectured long ago \cite{Susskind:1994sm} that the entangling surface in string theory acts like a brane on which open strings end.  Recently, we showed that this ``entanglement brane'' does indeed arise in a perturbative string calculation of entanglement entropy in a 2D string theory, dual to 2d Yang mills \cite{Donnelly:2016jet}.   In that work, we showed that the entanglement brane provides a geometrical description of the entanglement edge modes in 2D Yang Mills.   

Since Chern Simons theory is also dual to a (topological) string theory, it is natural to look for a string theory description of the entanglement edge modes in the form of a brane.    Indeed this was the original motivation for this work.   It has been known for sometime that the open string theory dual to U(N) Chern Simons theory on $S^3$ is the topological A-model on the deformed conifold $T^{*}S^3$, with $N$ branes wrapping $S^3$ \cite{Witten:1992fb} \cite{Gopakumar:1998ki}.  This open string theory is in turn dual to A model closed strings on the resolved conifold \cite{Gopakumar:1998ki}.  On this spacetime, the branes have been replaced by $N$ units of flux piercing the $S^3$.   However, by analogy with our work in 2D Yang Mills, it is natural to ask whether branes will reappear in the closed string A model due to the presence of an entangling surface.  We intend to pursue this question in the near future.

\section{Acknowledgements} 
We would like to thank Will Donnelly, Florian Hopfmuller, Janet Hung, Joan Simon, Rob Myers, Jeffrey Teo, Shinsei Ryu, and Rob Leigh for helpful discussions.  This work was supported in part by the DOE grant DE-SC0007894 and by Perimeter Institute for Theoretical Physics. Research at Perimeter Institute is supported by the Government of Canada through the Department of Innovation, Science and Economic Development Canada and by the Province of Ontario through the Ministry of Research, Innovation and Science.

\bibliographystyle{utphys}
\bibliography{CSedge}
\end{document}